\documentclass[prc,onecolumn,nofootinbib,showpacs,floatfix,a4paper]{revtex4}
\usepackage{enumerate}
\usepackage{booktabs}
\usepackage{color}
\usepackage{graphicx}
\usepackage{amsmath}
\usepackage{amssymb}
\usepackage{times}

\def\bvec#1{\mbox{\boldmath $#1$}}

\newcommand{\del}[2]{\frac{\partial #1}{\partial #2}}

\newcommand{\bra}{\langle}
\newcommand{\ket}{\rangle}

\newcommand{\idot}{\!\cdot\!}

\newcommand{\beq}{\begin{equation}}
\newcommand{\eeq}{\end{equation}}
\newcommand{\bea}{\begin{eqnarray}}
\newcommand{\eea}{\end{eqnarray}}

\def\fun#1#2{\lower3.6pt\vbox{\baselineskip0pt\lineskip.9pt
 \ialign{$\mathsurround=0pt#1\hfil##\hfil$\crcr#2\crcr\sim\crcr}}}

\begin{document}

\title{
Jost function formalism with complex potential
}

\author{K. Mizuyama$^{1,2}$, T. Dieu Thuy$^{3}$, N. Hoang Tung$^{4,5}$, D. Quang Tam$^{6}$,
  T. V. Nhan Hao$^{3,7}$}
\email{corresponding author: tvnhao@hueuni.edu.vn}

\affiliation{
  \textsuperscript{1} Institute of Research and Development, Duy Tan University, Danang 550000, Vietnam \\
  \textsuperscript{2} Faculty of Natural Sciences, Duy Tan University, Danang 550000, Vietnam \\
  \textsuperscript{3} Faculty of Physics, University of Education, Hue University, 34 Le Loi Street, Hue City, Vietnam \\
  \textsuperscript{4} Department of Nuclear Physics and Nuclear Engineering, Faculty of Physics an Engineering Physics, University of Science, Ho Chi Minh City, Vietnam \\
  \textsuperscript{5} Vietnam National University, Ho Chi Minh City, Vietnam \\
  \textsuperscript{6} Faculty of Basic Sciences, University of Medicine and Pharmacy, Hue University, Hue City 52000, Vietnam \\
  \textsuperscript{7} Center for Theoretical and Computational Physics, College of Education, Hue University, 34 Le Loi Street, Hue City, Vietnam
}

\date{\today}

\begin{abstract}
  The Jost function formalism is extended with use of the complex potential
  in this paper. We derive the Jost function by taking into account
  the dual state which is defined by the complex conjugate the complex
  Hamiltonian. By using the unitarity of the S-matrix which is defined by the
  Jost function, the optical theorem with the complex potential is also derived.
  The role of the imaginary part of the complex potential for both
  the bound states and the scattering states is figured out. The numerical
  calculation is performed by using the complex Woods-Saxon potential, and some
  numerical results are demonstrated to confirmed the properties of extended Jost
  function formalism.
\end{abstract}

\maketitle

\section{Introduction}

The optical model has been used to analyze and successfully reproduced the
experimental cross section of the nucleon-nucleus (NA) scattering at
higher energy region~\cite{optmodel}. As is well known, the optical potential is given
by the complex potential, and the imaginary part of the optical potential is
important for the quantitative reproduction of the experimental data because
it has the absorptive effect to the incident nucleon flux on the NA scattering
as written in many papers and textbooks. According to the Feshbach projection
theory~\cite{feshbach}, the origin of the complex optical potential is due to the
channels coupling at the intermediate states of the NA-scattering.
Nevertheless, the optical potential parameters are given phenomenological
so as to reproduce the experimental data~\cite{optpot,optpot2}.

Recently, the microscopic optical potential (MOP) based on the particle-vibration
coupling (PVC) method has been carried out. The elastic and inelastic
cross section of the NA-scattering have successfully reproduced the experimental data
without any fitting parameters because the MOP has been calculated by using
the effective nucleon-nucleon interaction
self-consistently~\cite{mizuyama,mizuyama2,hao, branchon}.
In the PVC method, the
MOP is represented by the Hartree-Fock mean field potential and self-energy function.
The self-energy function is given by the complex non-local function which represents
the coupling between the incoming nucleon and excitation of that target nucleus.
The excitation of the target is calculated by the self-consistent random phase
approximation (RPA). The elastic, inelastic and nucleon capture channels are
automatically taken into account in the self-energy function as the intermediate
states of the NA-scattering. The description of MOP by PVC is consistent with the
Feshbach projection theory.

The contribution of the optical potential can be interpreted as the mean contribution
from the coupling of many channels at higher incident incident energy of NA scattering.
At low energy, the basic shape of the cross section is characterized by the individual
contribution of shape resonances. As expected from the
R-matrix theory~\cite{rmatrix,rmatrix1,rmatrix2} and
results and analysis by continuum PVC~\cite{mizuyama}, the channel coupling effect will play
crucial role for the production of the sharp peaks of resonances. The absorptive effect
of optical potential becomes smaller. This is the fact which can be expected from the
absolute value of the reaction cross section (which is defined by the difference of the
total and elastic cross section).

The important role of the imaginary part of the complex potential is not only for the
nuclear reaction. The role of the complex potential for the nuclear structure has been
discussed.
In order to analyze the PVC effect on the single particle levels, the experimentally
determined spectroscopic factor has been analyzed by compering with the theoretically
calculated level density~\cite{cPVC}. It was confirmed that the experimental spectroscopic
factors are reasonably well reproduced in both senses; the position of the centroid
energies and the fragmentation. This is another aspect of the complex self-energy
function (complex optical potential)'s effect.

The Jost function is one of the useful tool to investigate the role of the potential
for bound states, resonances and scattering states, because it is possible to calculate
all of those states as the poles of the S-matrix on the complex energy/momentum plane
directly from the potential.
Recently, we extended the Jost function formalism based on the Hartree-Fock-Bogoliubov
(HFB) formalism in order to take into account the pairing~\cite{jost-hfb}.
This extension of the Jost
function is a kind of the extension to the multi-channel system because the HFB is also
a kind of two channel system in a broad sense. However, the potential in the Jost function
framework is still supposed to be real. Therefore, we shall extend the Jost function
framework with the complex potential in this paper.

\section{Jost function with the complex potential}

When the potential is given by the complex,
the hermiticity of the Hamiltonian is broken.
As is well-known, the dual Hilbert space which is defined by
the complex conjugate of the Hamiltonian $H^*$ exists
if the hermiticity of the Hamiltonian is broken.
In order to derive the Jost function with the complex potential,
it is necessary to take into account the dual space.

\subsection{Derivation of the Jost function}

In this section, we shall derive the Jost function with the
complex potential.

When the Schrodinger equation is given by
\begin{eqnarray}
  &&
  \left[
  -\frac{\hbar^2}{2m}
  \left(
  \del{^2}{r^2}
  -
  \frac{l(l+1)}{r^2}
  \right)
  +
  U_{lj}(r)
  \right]
  \varphi_{lj}(r;k)
  \nonumber\\
  &&
  =
  \epsilon(k)
  \varphi_{lj}(r;k),
  \label{eq01}
\end{eqnarray}
with the complex potential $U_{lj}(r)$,
the equation for the dual state is given by
\begin{eqnarray}
  &&
  \left[
  -\frac{\hbar^2}{2m}
  \left(
  \del{^2}{r^2}
  -
  \frac{l(l+1)}{r^2}
  \right)
  +
  U_{lj}^*(r)
  \right]
  \tilde{\varphi}_{lj}(r;k)
  \nonumber\\
  &&
  =
  \epsilon(k)
  \tilde{\varphi}_{lj}(r;k),
  \label{eq02}
\end{eqnarray}
where
\begin{eqnarray}
  \epsilon(k)
  =
  \frac{\hbar^2k^2}{2m}.
\end{eqnarray}

By using the Green's theorem, we can obtain the regular and irregular
solutions, $\varphi_{lj}^{(r)}(r;k)$ and $\varphi_{lj}^{(\pm)}(r;k)$ as
\begin{eqnarray}
  &&
  \varphi_{lj}^{(r)}(r;k)
  \nonumber\\
  &&
  =
  F_l(kr)
  \nonumber\\
  &&\hspace{10pt}
  +
  \int_0^\infty dr'
  g_{FR,l}(r,r';k)
  U_{lj}(r')
  \varphi_{lj}^{(r)}(r';k),
  \nonumber\\
  \label{LSvarr-compot1}
\end{eqnarray}
\begin{eqnarray}
  &&
  \tilde{\varphi}_{lj}^{(r)}(r;k)
  \nonumber\\
  &&
  =
  F_l(kr)
  \nonumber\\
  &&\hspace{10pt}
  +
  \int_0^\infty dr'
  g_{FR,l}(r,r';k)
  U^*_{lj}(r')
  \tilde{\varphi}_{lj}^{(r)}(r';k),
  \nonumber\\
  \label{LSvarr-compot2}
\end{eqnarray}
\begin{eqnarray}
  &&
  \varphi_{lj}^{(\pm)}(r;k)
  \nonumber\\
  &&
  =
  O^{(\pm)}_l(kr)
  \nonumber\\
  &&\hspace{10pt}
  +
  \int_0^\infty dr'
  g_{FL,l}(r,r';k)
  U_{lj}(r')
  \varphi_{lj}^{(\pm)}(r';k),
  \nonumber\\
  \label{LSvarpm-compot1}
\end{eqnarray}
and
\begin{eqnarray}
  &&
  \tilde{\varphi}_{lj}^{(\pm)}(r;k)
  \nonumber\\
  &&
  =
  O^{(\pm)}_l(kr)
  \nonumber\\
  &&\hspace{10pt}
  +
  \int_0^\infty dr'
  g_{FL,l}(r,r';k)
  U^*_{lj}(r')
  \tilde{\varphi}_{lj}^{(\pm)}(r';k),
  \nonumber\\
  \label{LSvarpm-compot2}
\end{eqnarray}
where $F_l(kr)=rj_l(kr)$ and $O_l^{(\pm)}=rh_l^{(\pm)}(kr)$, and
$g_{FR,l}(r,r';k)$ and $g_{FL,l}(r,r';k)$ are the Green's functions
defined by
\begin{eqnarray}
  g_{FR,l}(r,r';k)
  &\equiv&
  -
  \frac{2m}{\hbar^2}
  \frac{k}{2i}
  \theta(r-r')r,r'
  \nonumber\\
  &&\times
  \left[
    h^{(-)}_l(kr)h^{(+)}_l(kr')
    \right.
    \nonumber\\
    &&\left.
    \hspace{30pt}
    -
    h^{(-)}_l(kr')h^{(+)}_l(kr)
    \right],
  \label{gfrdef}
\end{eqnarray}
and
\begin{eqnarray}
  g_{FL,l}(r,r';k)
  &\equiv&
  \frac{2m}{\hbar^2}
  \frac{k}{2i}r,r'
  \theta(r'-r)
  \nonumber\\
  &&\hspace{10pt}
  \times
  \left[
    h^{(-)}_l(kr)h^{(+)}_l(kr')
    \right.
    \nonumber\\
    &&\hspace{50pt}
    \left.
    -
    h^{(-)}_l(kr')h^{(+)}_l(kr)
    \right].
  \label{gfldef}
\end{eqnarray}

By taking the limit of $r\to\infty$ in Eqs.(\ref{LSvarr-compot1})
and (\ref{LSvarr-compot2}), we can obtain
\begin{eqnarray}
  &&
  \lim_{r\to\infty}
  \varphi_{lj}^{(r)}(r;k)
  \nonumber\\
  &&
  \to
  F_l(kr)
  \nonumber\\
  &&\hspace{10pt}
  -
  \frac{2m}{\hbar^2}
  \frac{k}{2i}
  O^{(-)}_l(kr)
  \int_0^\infty dr'
  O^{(+)}_l(kr')
  U_{lj}(r')
  \varphi_{lj}^{(r)}(r';k)
  \nonumber\\
  &&\hspace{10pt}
  +
  \frac{2m}{\hbar^2}
  \frac{k}{2i}
  O^{(+)}_l(kr)
  \int_0^\infty dr'
  O^{(-)}_l(kr')
  U_{lj}(r')
  \varphi_{lj}^{(r)}(r';k),
  \nonumber\\
  \label{LSvarr-compot1-limit}
\end{eqnarray}
and
\begin{eqnarray}
  &&
  \lim_{r\to\infty}
  \tilde{\varphi}_{lj}^{(r)}(r;k)
  \nonumber\\
  &&
  =
  F_l(kr)
  \nonumber\\
  &&\hspace{10pt}
  -
  \frac{2m}{\hbar^2}
  \frac{k}{2i}
  O^{(-)}_l(kr)
  \int_0^\infty dr'
  O^{(+)}_l(kr')
  U^*_{lj}(r')
  \tilde{\varphi}_{lj}^{(r)}(r';k)
  \nonumber\\
  &&\hspace{10pt}
  +
  \frac{2m}{\hbar^2}
  \frac{k}{2i}
  O^{(+)}_l(kr)
  \int_0^\infty dr'
  O^{(-)}_l(kr')
  U^*_{lj}(r')
  \tilde{\varphi}_{lj}^{(r)}(r';k).
  \nonumber\\
  \label{LSvarr-compot2-limit}
\end{eqnarray}

Since the Jost function is defined as a coefficient
function to connect the regular and irregular
solution as
\begin{eqnarray}
  &&
  \varphi_{lj}^{(r)}(r;k)
  =
  \frac{1}{2}
  \left[
    J_{lj}^{(+)}(k)
    \varphi_{lj}^{(-)}(r;k)
    +
    J_{lj}^{(-)}(k)
    \varphi_{lj}^{(+)}(r;k)
    \right],
  \nonumber\\
  \label{defjost-compot1}
\end{eqnarray}
and
\begin{eqnarray}
  &&
  \tilde{\varphi}_{lj}^{(r)}(r;k)
  =
  \frac{1}{2}
  \left[
    \tilde{J}_{lj}^{(+)}(k)
    \tilde{\varphi}_{lj}^{(-)}(r;k)
    +
    \tilde{J}_{lj}^{(-)}(k)
    \tilde{\varphi}_{lj}^{(+)}(r;k)
    \right],
  \nonumber\\
  \label{defjost-compot2}
\end{eqnarray}
we can obtain
\begin{eqnarray}
  J_{lj}^{(\pm)}(k)
  =
  1
  \mp
  \frac{2m}{\hbar^2}
  \frac{k}{i}
  \int_0^\infty dr'
  O^{(\pm)}_l(kr')
  U_{lj}(r')
  \varphi_{lj}^{(r)}(r';k),
  \nonumber\\
  \label{jost-comp1}
\end{eqnarray}
and
\begin{eqnarray}
  \tilde{J}_{lj}^{(\pm)}(k)
  =
  1
  \mp
  \frac{2m}{\hbar^2}
  \frac{k}{i}
  \int_0^\infty dr'
  O^{(\pm)}_l(kr')
  U^*_{lj}(r')
  \tilde{\varphi}_{lj}^{(r)}(r';k).
  \nonumber\\
  \label{jost-comp2}
\end{eqnarray}
It is rather easy to prove the following expressions of the Jost function.
\begin{eqnarray}
  J_{lj}^{(\pm)}(k)
  =
  \pm
  \frac{2m}{\hbar^2}
  \frac{k}{i}
  W_{lj}(\varphi_{lj}^{(r)},\varphi_{lj}^{(\pm)};k),
  \label{jost-wron-cpot1}
\end{eqnarray}
and
\begin{eqnarray}
  \tilde{J}_{lj}^{(\pm)}(k)
  =
  \pm
  \frac{2m}{\hbar^2}
  \frac{k}{i}
  W_{lj}(\tilde{\varphi}_{lj}^{(r)},\tilde{\varphi}_{lj}^{(\pm)};k),
  \label{jost-wron-cpot2}
\end{eqnarray}
where $W_{lj}$ is the Wronskian is defined by
\begin{eqnarray}
  &&
  W_{lj}(\varphi_{lj}^{(r)},\varphi_{lj}^{(\pm)};k)
  \nonumber\\
  &&
  =
  \frac{\hbar^2}{2m}
  \left[
    \varphi_{lj}^{(r)}(r;k)\del{\varphi_{lj}^{(\pm)}(r;k)}{r}
    -
    \varphi_{lj}^{(\pm)}(r;k)\del{\varphi_{lj}^{(r)}(r;k)}{r}
    \right],
  \nonumber\\
  \label{wron-compot1}
\end{eqnarray}
and
\begin{eqnarray}
  &&
  W_{lj}(\tilde{\varphi}_{lj}^{(r)},\tilde{\varphi}_{lj}^{(\pm)};k)
  \nonumber\\
  &&
  =
  \frac{\hbar^2}{2m}
  \left[
    \tilde{\varphi}_{lj}^{(r)}(r;k)\del{\tilde{\varphi}_{lj}^{(\pm)}(r;k)}{r}
    -
    \tilde{\varphi}_{lj}^{(\pm)}(r;k)\del{\tilde{\varphi}_{lj}^{(r)}(r;k)}{r}
    \right].
  \nonumber\\
  \label{wron-compot2}
\end{eqnarray}

By taking the limit of $r\to 0$ for Eqs.(\ref{LSvarpm-compot1}) and
Eqs.(\ref{LSvarpm-compot2}), we can obtain
\begin{eqnarray}
  &&
  \lim_{r\to 0}
  \varphi_{lj}^{(\pm)}(r;k)
  \nonumber\\
  &&
  \to
  O_l^{(\pm)}(kr)
  \nonumber\\
  &&\hspace{10pt}
  +
  \frac{2m}{\hbar^2}
  \frac{k}{2i}
  O^{(-)}_l(kr)
  \int_0^\infty dr'
  O^{(+)}_l(kr')
  U_{lj}(r')
  \varphi_{lj}^{(\pm)}(r';k)
  \nonumber\\
  &&\hspace{10pt}
  -
  \frac{2m}{\hbar^2}
  \frac{k}{2i}
  O^{(+)}_l(kr)
  \int_0^\infty dr'
  O^{(-)}_l(kr')
  U_{lj}(r')
  \varphi_{lj}^{(\pm)}(r';k),
  \nonumber\\
  \label{LSvarpm-compot1-limit}
\end{eqnarray}
and
\begin{eqnarray}
  &&
  \lim_{r\to 0}
  \tilde{\varphi}_{lj}^{(\pm)}(r;k)
  \nonumber\\
  &&
  =
  O_l^{(\pm)}(kr)
  \nonumber\\
  &&\hspace{10pt}
  +
  \frac{2m}{\hbar^2}
  \frac{k}{2i}
  O^{(-)}_l(kr)
  \int_0^\infty dr'
  O^{(+)}_l(kr')
  U^*_{lj}(r')
  \tilde{\varphi}_{lj}^{(\pm)}(r';k)
  \nonumber\\
  &&\hspace{10pt}
  -
  \frac{2m}{\hbar^2}
  \frac{k}{2i}
  O^{(+)}_l(kr)
  \int_0^\infty dr'
  O^{(-)}_l(kr')
  U^*_{lj}(r')
  \tilde{\varphi}_{lj}^{(\pm)}(r';k),
  \nonumber\\
  \label{LSvarpm-compot2-limit}
\end{eqnarray}
Using
Eqs.(\ref{jost-wron-cpot1})-(\ref{jost-wron-cpot2}),
Eqs.(\ref{wron-compot1})-(\ref{wron-compot2}) and
Eqs.(\ref{LSvarpm-compot1-limit})-(\ref{LSvarpm-compot2-limit}),
we can obtain
\begin{eqnarray}
  J_{lj}^{(\pm)}(k)
  =
  1
  \mp
  \frac{2m}{\hbar^2}
  \frac{k}{i}
  \int_0^\infty dr'
  F_l(kr')
  U_{lj}(r')
  \varphi_{lj}^{(\pm)}(r';k),
  \nonumber\\
  \label{jost-comp3}
\end{eqnarray}
and
\begin{eqnarray}
  \tilde{J}_{lj}^{(\pm)}(k)
  =
  1
  \mp
  \frac{2m}{\hbar^2}
  \frac{k}{i}
  \int_0^\infty dr'
  F_l(kr')
  U^*_{lj}(r')
  \tilde{\varphi}_{lj}^{(\pm)}(r';k),
  \nonumber\\
  \label{jost-comp4}
\end{eqnarray}
Thus we obtained the Jost function by three types of expressions for each
$J_{lj}^{(\pm)}(k)$ and $\tilde{J}_{lj}^{(\pm)}(k)$ as given by
Eqs.(\ref{jost-comp1}), (\ref{jost-comp2}),
(\ref{jost-wron-cpot1}), (\ref{jost-wron-cpot2}),
(\ref{jost-comp3}) and (\ref{jost-comp4}).

The Jost functions satisfy the following symmetric properties.
\begin{eqnarray}
  J_{lj}^{(\pm)*}(k^*)
  =
  \tilde{J}_{lj}^{(\mp)}(k)
  \label{jost-prop-cpot1}
\end{eqnarray}
\begin{eqnarray}
  J_{lj}^{(\pm)}(-k)
  =
  J_{lj}^{(\mp)}(k),
  \label{jost-prop-cpot2}
\end{eqnarray}
and
\begin{eqnarray}
  \tilde{J}_{lj}^{(\pm)}(-k)
  =
  \tilde{J}_{lj}^{(\mp)}(k).
  \label{jost-prop-cpot3}
\end{eqnarray}

\subsection{Bound state}

The boundary condition for a bound state
$\epsilon_n=\epsilon(k_n)$ is given by
\begin{eqnarray}
  \left\{
  \begin{array}{cc}
    J_{lj}^{(+)}(k_n)=0 & \mbox{for $\varphi_{nlj}$} \\
    \tilde{J}_{lj}^{(+)}(\tilde{k}_n)=0 & \mbox{for $\tilde{\varphi}_{nlj}$}
  \end{array}
  \right.
\end{eqnarray}

From Eqs.(\ref{jost-prop-cpot1})-(\ref{jost-prop-cpot3}), we can
obtain
\begin{eqnarray}
  \tilde{J}_{lj}^{(+)}(k)
  =
  J_{lj}^{(+)*}(-k^*),
\end{eqnarray}
therefore we can find the relation between $k_n$ and $\tilde{k}_n$ as
\begin{eqnarray}
  \tilde{k}_n=-k_n^*.
  \label{bdcon}
\end{eqnarray}

The relation between the energy eigen value
for the dual state $\epsilon(\tilde{k}_n)$ and
$\epsilon(k_n)$ is given by
\begin{eqnarray}
  \epsilon(\tilde{k}_n)
  =
  \epsilon(-k^*_n)
  =
  \epsilon^*(k_n).
\end{eqnarray}
It is very easy to rove that the
orthogonality of the bound state wave function is given by
\begin{eqnarray}
  \int_0^\infty dr
  \tilde{\varphi}^*_{nlj}(r)
  \varphi_{mlj}(r)
  =\delta_{nm}.
  \label{orth-bd-cpot}
\end{eqnarray}
By dividing the potential into two parts, real and imaginary parts as
\begin{eqnarray}
  U_{lj}=U_{lj}^r-iU_{lj}^i,
\end{eqnarray}
and applying the two potential formula to the Green's function,
we can obtain the Dyson equation
\begin{eqnarray}
  &&
  G^{(\pm)}_{lj}(r,r';k)
  =
  G^{(\pm)}_{0,lj}(r,r';k)
  \nonumber\\
  &&
  +
  \int_0^\infty
  dr''
  G^{(\pm)}_{0,lj}(r,r'';k)
  (-iU_{lj}^i(r''))
  G^{(\pm)}_{lj}(r'',r';k),
  \nonumber\\
  \label{dyson-cpot-2pot}
\end{eqnarray}
where $G^{(\pm)}_{0,lj}(r,r';k)$ is the Green's function
which satisfies the following equation
\begin{eqnarray}
  \left(\epsilon(k)-h^0_{lj}(r)\right)
  G^{(\pm)}_{0,lj}(r,r';k)
  =
  \delta(r-r'),
\end{eqnarray}
with
\begin{eqnarray}
  h^0_{lj}(r)
  =
  -\frac{\hbar^2}{2m}
  \left(
  \del{^2}{r^2}
  -
  \frac{l(l+1)}{r^2}
  \right)
  +
  U^r_{lj}(r).
  \nonumber\\
\end{eqnarray}
Using the spectral representation of $G^{(\pm)}_{0,lj}(r,r';k)$
and $G^{(\pm)}_{lj}(r,r';k)$ in Eq.(\ref{dyson-cpot-2pot}),
we can obtain
\begin{eqnarray}
  \epsilon_{nlj}
  \sim
  \epsilon^0_{nlj}
  -i
  \frac{\bra\varphi_{nlj}^0|U^i_{lj}|\varphi_{nlj}\ket}%
       {\bra\varphi_{nlj}^0|\varphi_{nlj}\ket},%
       \label{approxres}
\end{eqnarray}
where $\varphi_{nlj}^0$ is the bound state wave function satisfies
$h_{lj}^0|\varphi_{nlj}^0\ket=\epsilon_{nlj}^0|\varphi_{nlj}^0\ket$.

If the potential is real function ($U_{lj}^*\to U_{lj}$),
obviously $\tilde{J}_{lj}^{(+)}(k)\to J_{lj}^{(+)}(k)$,
$\tilde{k}_n\to k_n$. However,
\begin{eqnarray}
  k_n=-k_n^*,
\end{eqnarray}
is required due to Eq.(\ref{bdcon}).
Therefore, we can find that $k_n$ for the real potential
becomes pure imaginary, and the energy eigen value becomes real
number.

\subsection{Scattering state and optical theorem}

The scattering wave function $\psi^{(+)}_{lj}(r;k)$
has the asymptotic behavior
\begin{eqnarray}
  \lim_{r\to\infty}
  \psi^{(+)}_{lj}(r;k)
  \to
  \frac{1}{2}
  \left[
    O_l^{(-)}(kr)
    +
    S_{lj}(k)
    O_l^{(+)}(kr),
    \right]
  \nonumber\\
  \label{scat-cpot1}
\end{eqnarray}
with the S-matrix $S_{lj}(k)$. The scattering wave function
as the dual state of Eq.(\ref{scat-cpot1}) has the asymptotic behavior
\begin{eqnarray}
  \lim_{r\to\infty}
  \tilde{\psi}^{(+)}_{lj}(r;k)
  \to
  \frac{1}{2}
  \left[
    O_l^{(-)}(kr)
    +
    \tilde{S}_{lj}(k)
    O_l^{(+)}(kr)
    \right],
  \nonumber\\
  \label{scat-cpot2}
\end{eqnarray}
with the S-matrix $\tilde{S}_{lj}(k)$.
By concerning the asymptotic behavior of
Eqs.(\ref{LSvarr-compot1}) and (\ref{LSvarr-compot2}),
we can find that $\psi^{(+)}_{lj}(r;k)$ and
$\tilde{\psi}^{(+)}_{lj}(r;k)$ are defined by
\begin{eqnarray}
  \psi^{(+)}_{lj}(r;k)
  =
  \frac{\varphi_{lj}^{(r)}(r;k)}{J_{lj}^{(+)}(k)},
  \label{scat-cpot3}
\end{eqnarray}
and
\begin{eqnarray}
  \tilde{\psi}^{(+)}_{lj}(r;k)
  =
  \frac{\tilde{\varphi}_{lj}^{(r)}(r;k)}{\tilde{J}_{lj}^{(+)}(k)}.
  \label{scat-cpot4}
\end{eqnarray}
The S-matrix $S_{lj}(k)$ and $\tilde{S}_{lj}(k)$ are defined by
\begin{eqnarray}
  S_{lj}(k)
  =
  \frac{J_{lj}^{(-)}(k)}{J_{lj}^{(+)}(k)},
  \label{smat-cpot1}
\end{eqnarray}
\begin{eqnarray}
  \tilde{S}_{lj}(k)
  =
  \frac{\tilde{J}_{lj}^{(-)}(k)}{\tilde{J}_{lj}^{(+)}(k)},
  \label{smat-cpot2}
\end{eqnarray}
respectively.

From Eqs.(\ref{jost-prop-cpot1})-(\ref{jost-prop-cpot3}),
we can derive the following properties of the S-matrix.
\begin{eqnarray}
  S^*_{lj}(k^*)
  =
  \tilde{S}_{lj}^{-1}(k),
  \label{smat-cpot-prop1}
\end{eqnarray}
\begin{eqnarray}
  \tilde{S}^*_{lj}(k^*)
  =
  S_{lj}^{-1}(k)
  \label{smat-cpot-prop2},
\end{eqnarray}
therefore we can obtain
\begin{eqnarray}
  S^*_{lj}(k^*)\tilde{S}_{lj}(k)
  =
  \tilde{S}^*_{lj}(k^*)S_{lj}(k)
  =
  1.
  \label{Unitarity0}
\end{eqnarray}
The scattering states (continuum) is defined
on the real axis of the complex momentum plane.
Therefore the Unitarity of the S-matrix in the
complex potential system is given by
\begin{eqnarray}
  S^*_{lj}(k)\tilde{S}_{lj}(k)
  =
  \tilde{S}^*_{lj}(k)S_{lj}(k)
  =
  1,
  \label{Unitarity1}
\end{eqnarray}
when $k^*=k$ ({\it i.e. }$k$ is real).

The T-matrix is defined by
\begin{eqnarray}
  T_{lj}(k)
  &=&
  \frac{i}{2}
  \left(
  S_{lj}(k)-1
  \right)
  \label{tmat1-cpot}
  \\
  &=&
  \frac{2mk}{\hbar^2}
  \int_0^\infty dr
  F_l(kr)
  U_{lj}(r)
  \psi_{lj}^{(+)}(r;k),
  \nonumber\\
  \label{tmat2-cpot}
\end{eqnarray}
and
\begin{eqnarray}
  \tilde{T}_{lj}(k)
  &=&
  \frac{i}{2}
  \left(
  \tilde{S}_{lj}(k)-1
  \right)
  \label{tmat3-cpot}
  \\
  &=&
  \frac{2mk}{\hbar^2}
  \int_0^\infty dr
  F_l(kr)
  U^*_{lj}(r)
  \tilde{\psi}_{lj}^{(+)}(r;k).
  \nonumber\\
  \label{tmat4-cpot}
\end{eqnarray}

From Eqs.(\ref{tmat2-cpot}) and (\ref{tmat4-cpot}),
we can obtain
\begin{eqnarray}
  T^*_{lj}(k^*)
  &=&
  S^*_{lj}(k^*)
  \tilde{T}_{lj}(k),
  \label{tmat5-cpot}
\end{eqnarray}
and
\begin{eqnarray}
  \tilde{T}^*_{lj}(k^*)
  &=&
  \tilde{S}^*_{lj}(k^*)
  T_{lj}(k),
  \label{tmat6-cpot}
\end{eqnarray}
Eq.(\ref{tmat5-cpot}) can be rewritten as
\begin{eqnarray}
  -
  \frac{1}{2i}
  \left(
  \tilde{T}_{lj}(k)
  -
  T^*_{lj}(k^*)
  \right)
  &=&
  T^*_{lj}(k^*)\tilde{T}_{lj}(k).
  \nonumber\\
  \label{tmat7-cpot}
\end{eqnarray}
This is the generalized optical theorem.

When the potential is real, $\tilde{T}_{lj}(k)=T_{lj}(k)$.
Therefore Eq.(\ref{tmat7-cpot}) becomes
\begin{eqnarray}
  -
  \mbox{Im }
  T_{lj}(k)
  &=&
  |T_{lj}(k)|^2,
  \nonumber\\
  \label{tmat8-cpot}
\end{eqnarray}
on the real axis of the complex momentum $k$ ({\it i.e. } $k^*=k$).

The Green's theorem leads the following Lippmann-Schwinger equations
\begin{eqnarray}
  &&
  \psi_{lj}^{(+)}(r;k)
  =
  F_l(kr)
  \nonumber\\
  &&
  +
  \int_0^\infty dr'
  G^{(+)}_{F,l}(r,r';k)
  U_{lj}(r')
  \psi_{lj}^{(+)}(r';k),
  \nonumber\\
  \label{LSeq-cpot1}
\end{eqnarray}
and
\begin{eqnarray}
  &&
  \tilde{\psi}_{lj}^{(+)}(r;k)
  =
  F_l(kr)
  \nonumber\\
  &&
  +
  \int_0^\infty dr'
  G^{(+)}_{F,l}(r,r';k)
  U^*_{lj}(r')
  \tilde{\psi}_{lj}^{(+)}(r';k),
  \nonumber\\
  \label{LSeq-cpot2}
\end{eqnarray}
using the Green's function defined by
\begin{eqnarray}
  &&
  G^{(\pm)}_{F,l}(r,r';k)
  \nonumber\\
  &&=
  \mp i\frac{2mk}{\hbar^2}
  \nonumber\\
  &&\times
  \left[
    \theta(r-r')
    F_l(kr')
    O_l^{(\pm)}(kr)
    \right.
    \nonumber\\
    &&\hspace{20pt}
    +
    \left.
    \theta(r'-r)
    F_l(kr)
    O_l^{(\pm)}(kr')
    \right].
  \nonumber\\
  \label{GFdef-cpot}
\end{eqnarray}
This is the Green's function which is written in many textbook
as the free particle Green's function.

\begin{figure}[htbp]
\includegraphics[scale=0.35,angle=-90]{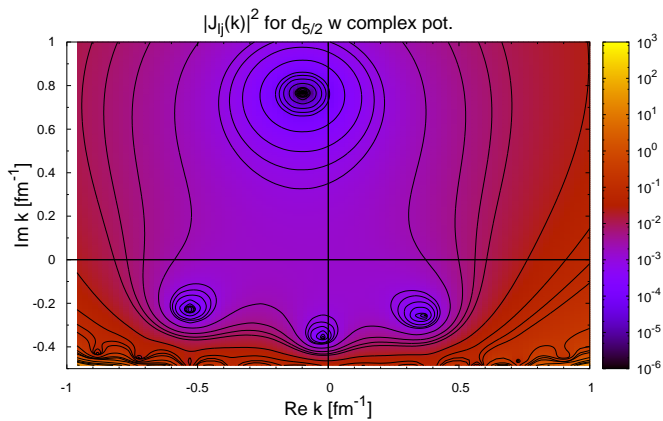}
\caption{(Color online) The square of the Jost function $|J_{lj}^{(+)}(k)|^2$ for $d_{5/2}$
  which is plotted as a function of the complex momentum $k$.}
\label{fig1}
\end{figure}
\begin{figure}[htbp]
\includegraphics[scale=0.35,angle=-90]{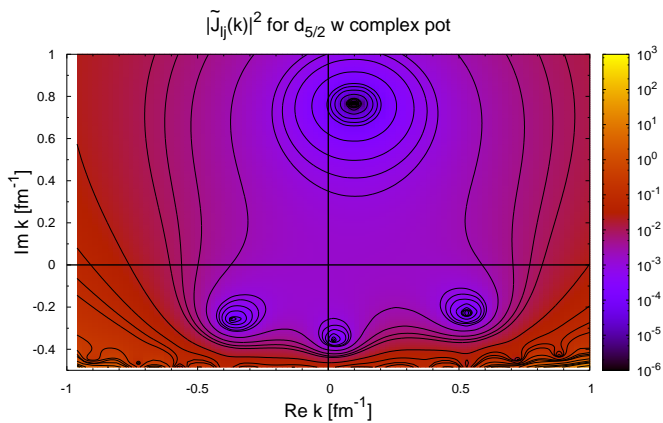}
\caption{(Color online) The same as Fig.\ref{fig1} but for $|\tilde{J}_{lj}^{(+)}(k)|^2$.}
\label{fig2}
\end{figure}

Also, the Green's theorem leads the another types of equations;
\begin{eqnarray}
  \psi_{lj}^{(+)}(r;k)
  =
  F_l(kr)
  +
  \int_0^\infty dr'
  G^{(+)}_{lj}(r,r';k)
  U_{lj}(r')
  F_l(kr'),
  \nonumber\\
  \label{LSeq-cpot3}
\end{eqnarray}
and
\begin{eqnarray}
  \tilde{\psi}_{lj}^{(+)}(r;k)
  =
  F_l(kr)
  +
  \int_0^\infty dr'
  \tilde{G}^{(+)}_{lj}(r,r';k)
  U^*_{lj}(r')
  F_l(kr'),
  \nonumber\\
  \label{LSeq-cpot4}
\end{eqnarray}
using the Green's function defined by
\begin{eqnarray}
  &&
  G^{(\pm)}_{lj}(r,r';k)
  \nonumber\\
  &&=
  \mp i\frac{2m}{\hbar^2}
  \frac{k}{J_{lj}^{(\pm)}(k)}
  \nonumber\\
  &&\times
  \left[
    \theta(r-r')
    \varphi_{lj}^{(r)}(r';k)
    \varphi_{lj}^{(\pm)}(r;k)
    \right.
    \nonumber\\
    &&\hspace{20pt}
    +
    \left.
    \theta(r'-r)
    \varphi_{lj}^{(r)}(r;k)
    \varphi_{lj}^{(\pm)}(r';k)
    \right],
  \nonumber\\
  \label{Gdef-cpot1}
\end{eqnarray}
and
\begin{eqnarray}
  &&
  \tilde{G}^{(\pm)}_{lj}(r,r';k)
  \nonumber\\
  &&=
  \mp i\frac{2m}{\hbar^2}
  \frac{k}{\tilde{J}_{lj}^{(\pm)}(k)}
  \nonumber\\
  &&\times
  \left[
    \theta(r-r')
    \tilde{\varphi}_{lj}^{(r)}(r';k)
    \tilde{\varphi}_{lj}^{(\pm)}(r;k)
    \right.
    \nonumber\\
    &&\hspace{20pt}
    +
    \left.
    \theta(r'-r)
    \tilde{\varphi}_{lj}^{(r)}(r;k)
    \tilde{\varphi}_{lj}^{(\pm)}(r';k)
    \right],
  \nonumber\\
  \label{Gdef-cpot2}
\end{eqnarray}
$\psi_{lj}^{(-)}(r;k)$ and $\tilde{\psi}_{lj}^{(-)}(r;k)$ are defined by
\begin{eqnarray}
  \psi_{lj}^{(-)}(r;k)
  \equiv
  \psi_{lj}^{(+)*}(r;k^*),
\end{eqnarray}
and
\begin{eqnarray}
  \tilde{\psi}_{lj}^{(-)}(r;k)
  \equiv
  \tilde{\psi}_{lj}^{(+)*}(r;k^*),
\end{eqnarray}
because $\psi_{lj}^{(-)}(r;k)$ and $\tilde{\psi}_{lj}^{(-)}(r;k)$ are
the time-reversal state of
$\psi_{lj}^{(+)}(r;k)$ and $\tilde{\psi}_{lj}^{(+)}(r;k)$.
Therefore we can obtain
\begin{eqnarray}
  &&
  \psi_{lj}^{(-)}(r;k)
  \nonumber\\
  &&
  =
  F_l(kr)
  +
  \int_0^\infty dr'
  G^{(-)}_{F,l}(r,r';k)
  U^*_{lj}(r')
  \psi_{lj}^{(-)}(r';k)
  \nonumber\\
  \label{LSeq-cpot5}
  \\
  &&
  =
  F_l(kr)
  +
  \int_0^\infty dr'
  \tilde{G}^{(-)}_{lj}(r,r';k)
  U^*_{lj}(r')
  F_l(kr')
  \label{LSeq-cpot6}
  \\
  &&
  =
  \frac{\tilde{\varphi}_{lj}^{(r)}(r;k)}{\tilde{J}_{lj}^{(-)}(k)},
  \label{LSeq-cpot7}
\end{eqnarray}
and
\begin{eqnarray}
  &&
  \tilde{\psi}_{lj}^{(-)}(r;k)
  \nonumber\\
  &&
  =
  F_l(kr)
  +
  \int_0^\infty dr'
  G^{(-)}_{F,l}(r,r';k)
  U_{lj}(r')
  \tilde{\psi}_{lj}^{(-)}(r';k)
  \nonumber\\
  \label{LSeq-cpot8}
  \\
  &&
  =
  F_l(kr)
  +
  \int_0^\infty dr'
  G^{(-)}_{lj}(r,r';k)
  U_{lj}(r')
  F_l(kr')
  \label{LSeq-cpot9}
  \\
  &&
  =
  \frac{\varphi_{lj}^{(r)}(r;k)}{J_{lj}^{(-)}(k)},
  \label{LSeq-cpot10}
\end{eqnarray}
From Eqs.(\ref{LSeq-cpot1}), (\ref{LSeq-cpot2}),
(\ref{LSeq-cpot3}), (\ref{LSeq-cpot4}),
(\ref{LSeq-cpot5}), (\ref{LSeq-cpot6}),
(\ref{LSeq-cpot8}) and (\ref{LSeq-cpot9}),
we can derive the following Dyson equations
\begin{eqnarray}
  &&
  G^{(\pm)}_{lj}(r,r';k)
  \nonumber\\
  &&
  =
  G^{(\pm)}_{F,l}(r,r';k)
  \nonumber\\
  &&
  +
  \int_0^\infty dr''
  G^{(\pm)}_{F,l}(r,r'';k)
  U_{lj}(r'')
  G^{(\pm)}_{lj}(r'',r';k),
  \nonumber\\
  \label{dyson-cpot1}
\end{eqnarray}
and
\begin{eqnarray}
  &&
  \tilde{G}^{(\pm)}_{lj}(r,r';k)
  \nonumber\\
  &&
  =
  G^{(\pm)}_{F,l}(r,r';k)
  \nonumber\\
  &&
  +
  \int_0^\infty dr''
  G^{(\pm)}_{F,l}(r,r'';k)
  U^*_{lj}(r'')
  \tilde{G}^{(\pm)}_{lj}(r'',r';k).
  \nonumber\\
  \label{dyson-cpot2}
\end{eqnarray}
Using Eqs.(\ref{LSeq-cpot1}), (\ref{LSeq-cpot2}) and (\ref{dyson-cpot2}),
we can derive
\begin{eqnarray}
  &&
  \psi_{lj}^{(+)}(r;k)
  \nonumber\\
  &&
  =
  \tilde{\psi}_{lj}^{(+)}(r;k)
  \nonumber\\
  &&
  +
  \int_0^\infty dr'
  \tilde{G}^{(+)}_{lj}(r,r';k)
  \left(U_{lj}(r')-U^*_{lj}(r')\right)
  \psi_{lj}^{(+)}(r';k).
  \nonumber\\
  \label{psitildepsi}
\end{eqnarray}
It should be noted that
$\tilde{\psi}_{lj}^{(+)}(r;k)=\psi_{lj}^{(+)}(r;k)$
when the potential is real ({\it i.e.} $U_{lj}=U^*_{lj}$).

From the limit $r\to\infty$ of Eq.(\ref{psitildepsi}),
we can obtain
\begin{eqnarray}
  &&
  \tilde{T}_{lj}(k)
  \nonumber\\
  &&
  =
  T_{lj}(k)
  \nonumber\\
  &&
  -
  \frac{2mk}{\hbar^2}
  \int_0^\infty dr'
  \tilde{\psi}_{lj}^{(+)}(r';k)
  \left(U_{lj}(r')-U^*_{lj}(r')\right)
  \psi_{lj}^{(+)}(r';k).
  \nonumber\\
  \label{psitildepsi2}
\end{eqnarray}

By inserting Eq.(\ref{psitildepsi2}) into Eq.(\ref{tmat7-cpot}),
the generalized optical theorem is rewritten as
\begin{eqnarray}
  &&
  -
  \mbox{Im }
  T_{lj}(k)
  \nonumber\\
  &&
  =
  |T_{lj}(k)|^2
  -
  \frac{2mk}{\hbar^2}
  \int_0^\infty dr'
  |\psi_{lj}^{(+)}(r';k)|^2
  \mbox{ Im }U_{lj}(r'),
  \nonumber\\
  \label{opttheorem-cpot}
\end{eqnarray}
when $k$ is real.

This is the optical theorem with the complex potential.
The left hand side of Eq.(\ref{opttheorem-cpot})
represents the total cross section.
In the right hand side of Eq.(\ref{opttheorem-cpot}),
the 1st term represents the elastic scattering
cross section, the 2nd term represents
the difference between the total and elastic cross
section. When $\mbox{ Im }U_{lj}(r)$ is negative,
the potential is absorptive because the 2nd term is
positive.
(The signature of the absorption by potential is
given by $\sigma_{tot}>\sigma_{el}$.)

\begin{figure}[htbp]
\includegraphics[scale=0.35,angle=-90]{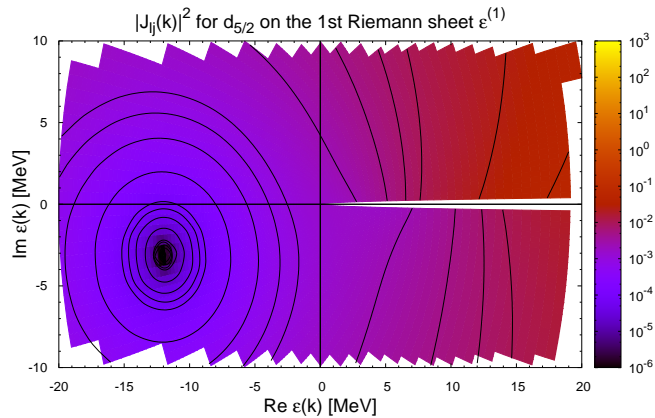}
\caption{(Color online) The square of the Jost function $|J_{lj}^{(+)}(k)|^2$ for $d_{5/2}$
  which is plotted on the first Riemann sheet $\epsilon^{(1)}$.}
\label{fig3}
\end{figure}
\begin{figure}[htbp]
\includegraphics[scale=0.35,angle=-90]{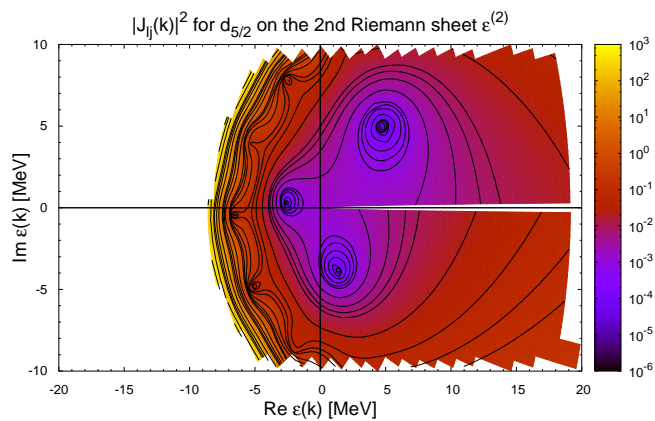}
\caption{(Color online) The same as Fig.\ref{fig3} but on the second Riemann sheet $\epsilon^{(2)}$.}
\label{fig4}
\end{figure}

\section{Numerical results}

We adopt the complex Woods-Saxon potential which is given by
\begin{eqnarray}
  U_{lj}(r)
  &=&
  -\left[
    V_1f_{ws}(x_1)
    +
    V_{2}\frac{1}{a_2r}g_{ws}(x_2)
    \bvec{l}\idot\bvec{\sigma}
    \right]
  \nonumber\\
  &&-i\left[
    V_3f_{ws}(x_3)
    +V_{4}\frac{1}{a_4r}g_{ws}(x_4)
    \bvec{l}\idot\bvec{\sigma}
    \right]
  \label{wspot}
  \\
  &&
  f_{ws}(x)
  =
  \frac{1}{1+e^x},
  \hspace{5pt}
  g_{ws}(x)
  =
  -\frac{df_{ws}(x)}{dx}
  \\
  &&
  x_i=(r-R_i)/a_i.
  \\
  &&
  R_i=\alpha_iA^{\gamma_i}+\beta_i,
  \hspace{5pt}
  (i\in 1,\cdots,4)
  \\
  \bvec{l}\idot\bvec{\sigma}
  &=&
  j(j+1)-l(l+1)-\frac{3}{4}.
\end{eqnarray}
As the original set of the parameters, we adopt
\begin{eqnarray*}
  &&A=24,\\
  &&V_1=51.0 \mbox{ MeV}, \hspace{10pt}V_2=17.0 \mbox{ MeV fm$^2$},\\
  &&V_3=5.0 \mbox{ MeV}, \hspace{10pt}V_4=0.0 \mbox{ MeV fm$^2$}, \\
  &&a_1=\cdots=a_4=0.67 \mbox{ fm }, \\
  &&\alpha_1=\cdots=\alpha_4=1.27 \mbox{ fm }, \\
  &&\beta_1=\cdots=\beta_4=0.0 \mbox{ fm }, \\
  &&\gamma_1=\cdots=\gamma_4=0.33 \mbox{ fm }.
\end{eqnarray*}

\begin{figure}[htbp]
\includegraphics[scale=0.35,angle=-90]{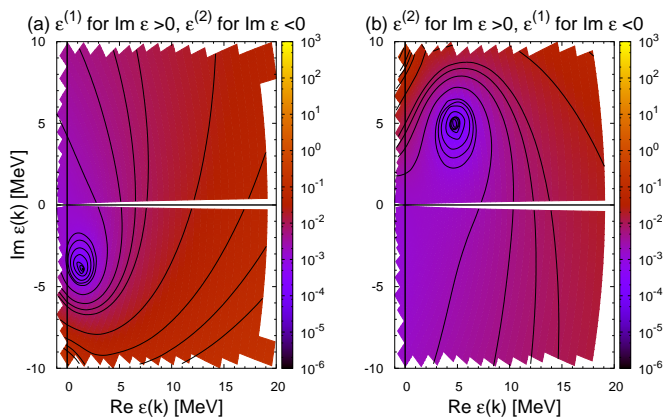}
\caption{(Color online) The analytic continuation between the first and second Riemann sheets.
  In the left panel, the upper-half of Fig.\ref{fig3} and the lower-half of
  Fig.\ref{fig4} are shown.
  In the right panel, the upper-half of Fig.\ref{fig4} and the lower-half of
  Fig.\ref{fig3} are shown.}
\label{fig5}
\end{figure}


In Figs.\ref{fig1} and \ref{fig2}, the numerical results of $|J_{lj}^{(+)}(k)|^2$
and $|\tilde{J}_{lj}^{(+)}(k)|^2$ for $d_{5/2}$ calculated with the complex Woods-Saxon
potential are shown on the complex momentum-$k$ plane. Because of the definition of
the S-matrix Eqs.(\ref{smat-cpot1}) and (\ref{smat-cpot2}), the minimum points which
correspond to $|J_{lj}^{(+)}(k)|^2=0$ in Fig.\ref{fig1} and $|\tilde{J}_{lj}^{(+)}(k)|^2=0$
in Fig.\ref{fig2} represent the poles of $S_{lj}(k)$ and $\tilde{S}_{lj}(k)$, respectively.
We can see the symmetric properties between $J_{lj}^{(+)}(k)$ and $\tilde{J}_{lj}^{(+)}(k)$
which are given by Eqs.(\ref{jost-prop-cpot1})-(\ref{jost-prop-cpot3}) in comparison
between Fig.\ref{fig1} and Fig.\ref{fig2}.
It should be noted that the S-matrix poles will be found symmetrically on the Im $k$
axis, and  the bound states appear on the Im $k$ axis, if the potential is given by
real function.

Since the energy is represented by the momentum as $\epsilon(k)=\frac{\hbar^2k^2}{2m}$,
two kinds of the Riemann sheets of the complex energy, the first and second Riemann sheets
($\epsilon^{(1)}$ and $\epsilon^{(2)}$), are defined by $k$ for Im $k >0$ and
Im $k <0$, respectively.

In Figs.\ref{fig3} and \ref{fig4}, $|J_{lj}^{(+)}(k)|^2$ for $d_{5/2}$ is represented
on $\epsilon^{(1)}$- and $\epsilon^{(2)}$-planes, respectively. Note that
Figs.\ref{fig3} and \ref{fig4} are corresponding to the upper-half and lower-half of
the complex-$k$ plane shown in Fig.\ref{fig1}, respectively.
One can see the discontinuity between the first and fourth quadrant across the
branch-cut which is defined by the real axis of the complex energy plane in the
positive region in both Figs.\ref{fig3} and \ref{fig4}.

The analytic continuation of $\epsilon^{(1)}$- and $\epsilon^{(2)}$-plane is shown in
Fig.\ref{fig5}. The first quadrant of Fig.\ref{fig3} is connected with the fourth
quadrant of Fig.\ref{fig4}, and the first quadrant of Fig.\ref{fig4} is connected with
the fourth quadrant of Fig.\ref{fig3}. This is due to the regularity of the Jost
function on the complex momentum plane.

In Table.\ref{splevels}, we show the numerical results for the single particle levels
with the real potential and complex potential obtained by searching the zeros of the
Jost function on the complex momentum plane.
The imaginary part of the single particle
levels Im $e_n$ obtained by using the complex potential are given by negative values.
This is consistent with the approximated formula given by Eq.(\ref{approxres}) since
the imaginary part of the potential is given by the negative value in this study.
The real part of the single particle levels are slightly shifted to higher energy due
to the effect of the imaginary part of the potential.

\begin{table}[htbp]
  \caption{Single particle levels for the bound neutrons for
    the real potential and complex potential, respectively.}
  \label{splevels}
  \centering
  \begin{tabular}{ccccccc}
    \\
    \hline
    & & & \multicolumn{2}{c}{Real potential} & \multicolumn{2}{c}{Complex potential}\\
    & & & \multicolumn{2}{c}{($V_3=0$ MeV)}   & \multicolumn{2}{c}{($V_3=5.0$ MeV)}  \\
    \hline
    $n$ & $l$ & $2j$ & Re $e_n$  & Im $e_n$ & Re $e_n$  & Im $e_n$\\
    \hline
    1 &  0 &  1 &  -34.7800  &   0.0000 & -34.7596  &  -4.3341 \\
    1 &  1 &  3 &  -23.5408  &   0.0000 & -23.4992  &  -3.7736 \\
    1 &  1 &  1 &  -19.7961  &   0.0000 & -19.7520  &  -3.8138 \\
    1 &  2 &  5 &  -12.0097  &   0.0000 & -11.9410  &  -3.1081 \\
    2 &  0 &  1 &   -8.6164  &   0.0000 &  -8.5116  &  -2.7687 \\
    1 &  2 &  3 &   -5.3174  &   0.0000 &  -5.2062  &  -2.9906 \\
    1 &  3 &  7 &   -0.7096  &   0.0000 &  -0.5793  &  -2.2661 \\
    \hline
  \end{tabular}
\end{table}

\section{Summary}

In this study, we extended the Jost function formalism based on the complex potential.
Since the Jost function is defined as the coefficient function to connect the regular
and irregular solutions of the Schrodinger equation, the Jost function can be derived
by finding the relation between the regular and irregular solutions by using the Green's
theorem with the proper boundary conditions for each solutions. In the system defined by
the Hamiltonian $H$ with the complex potential, two kinds of the Jost function are
defined, the Jost function for the system defined by $H$ and the Jost function for the dual
system defined by $H^*$. In order to make sure our derivation, we derived the symmetric
properties of the Jost function and confirmed them by the numerical results represented
on the complex energy/momentum plane.

As is written in many textbooks, the generalized optical theorem which includes the
absorption as the effect of the imaginary part of the complex potential is rather well
known. In order to confirm our derivation of the Jost function, firstly we derived the
generalized unitarity of the S-matrix by using the Jost function which is calculated
by the complex potential. Using the generalized unitarity of the S-matrix, we derived
the generalized optical theorem.
In order to confirm the effect of the imaginary part of the complex potential to the
bound states, we derived an approximated formula which shows the effect of the imaginary
part of the potential to the single particle levels by using the Green's function method.
And we confirmed that the numerical results for the single particle levels obtained by
using the Jost function are consistent with the derived formula.

The results of Table~\ref{splevels} are not qualitatively consistent with the results
of the previous cPVC calculation.
According to the discussion and results in \cite{cPVC}, the complex potential which is
calculated as the self-energy function within the PVC has the effect to shift the single
particle levels to lower energy, and provides large fragmentation to the single particle
levels far from the Fermi level. As shown in \cite{mizuyama,mizuyama2,hao, branchon},
the PVC self-energy function works
very well as the microscopic optical potential also for the description of the
$NA$-scattering cross section.

For the quantitative reproduction of the experimental data of the $NA$-scattering,
the global optical potential has adopted the complex Woods-Saxon form with the
energy dependence. The energy dependence has been given by adjusting the experimental
data, but the energy dependence has been given only for the positive energy region,
except the Dispersive Optical Potential~\cite{dspopt}.
In order to obtain the proper interpretation of physics from the analysis of the
experimental data using the phenomenological optical potential, the optical potential
should be available for both the nuclear structure (single particle levels and their
fragmentation and so on) and the nuclear reaction. The Jost function may be a powerful
tool to construct such a new type of the optical potential.

According to the Feshbach projection theory~\cite{feshbach},
the origin of the complex potential
is the coupling of channels, and the channel-coupling equation can be reduced to the
single channel problem by introducing the complex optical potential. As we showed in
this paper, two kinds of the Riemann sheets are defined for two types of the Jost
function with the complex potential.
On the other hand, the multiple Riemann sheets (more than two, depending on the number
of channels) are expected for the channel-coupling equation (The HFB framework is also
a kind of the channel-coupling method of two channels in a broad sense, and three types of
the Riemann sheets are defined within the HFB framework~\cite{jost-hfb}).
Clarifying the relationship between the Riemann surface defined by the complex optical
potential and the coupled-channel method is also a very interesting subject and one of
the directions for future research using the method of Jost functions.

\section{Acknowledgments}
This work is funded by Vietnam National Foundation for Science and Technology Development (NAFOSTED)
under grant number “103.04-2019.329”.

\end{document}